\documentclass[aps,prb,twocolumn,superscriptaddress,floatfix,longbibliography,citeautoscript]{revtex4-2}

\usepackage{amsmath,amssymb} 
\usepackage{bm} 
\usepackage{graphicx} 
\usepackage{comment} 
\usepackage[normalem]{ulem} 

\usepackage{xcolor} 
\definecolor{lightgray}{gray}{0.6}
\definecolor{medgray}{gray}{0.4}
\definecolor{mRed}{RGB}{230, 0, 50}
\colorlet{newtextColor}{mRed}

\newif\ifptitle
\newif\ifpnumber
\newcounter{para}
\newcommand\ptitle[1]{\par\refstepcounter{para}
{\ifpnumber{\noindent\textcolor{lightgray}{\textbf{\thepara}}\indent}\fi}
{\ifptitle{\textbf{[{#1}]}}\fi}}

\usepackage{hyperref}
\hypersetup{
colorlinks=true,
urlcolor= blue,
citecolor=blue,
linkcolor= blue,
}

\newcommand{\Vs}{\ensuremath{V_{\mathrm{s}}}}
\newcommand{\Is}{\ensuremath{I_{\mathrm{s}}}}
\newcommand{\Ve}{\ensuremath{V_{\mathrm{exc}}}}

\newcommand{\heng}{School of Engineering \& Applied Sciences, Harvard University, Cambridge, MA 02138, USA}
\newcommand{\hphys}{Department of Physics, Harvard University, Cambridge, MA 02138, USA}
\newcommand{\ORNL} {Materials Science \& Technology Division, Physical Sciences Directorate,
Oak Ridge National Lab, Oak Ridge, TN, 37831, USA
}

\ptitlefalse
\pnumberfalse

\begin{document}

\title{Nanoscale electronic variations in altermagnetic \texorpdfstring{$\alpha$}{alpha}-MnTe}

\author{Zeyu Ma}
\thanks{Equal contribution}
\affiliation{\heng}
\author{Yidi Wang}
\thanks{Equal contribution}
\affiliation{\hphys}
\author{Gal Tuvia}
\thanks{Equal contribution}
\email[]{galtuvia@g.harvard.edu}
\affiliation{\hphys}
\author{Kevin Hauser}
\affiliation{\hphys}
\author{Jiaqiang Yan}
\affiliation{\ORNL}
\author{Jennifer E. Hoffman}
\email[]{jhoffman@physics.harvard.edu}
\affiliation{\heng}
\affiliation{\hphys}
\date{\today}

\begin{abstract}
Altermagnets exhibit spin-split electronic bands like ferromagnets, yet they are magnetically compensated like collinear antiferromagnets. Altermagnets thus combine the benefits of ferromagnets and antiferromagnets, opening routes to tailored materials and applications. However, reported bulk signatures such as the anomalous Hall response in candidate altermagnets have been inconsistent across samples, suggesting that inhomogeneity may affect their functionality in electronic devices. Here, we use low-temperature scanning tunneling microscopy and spectroscopy to map the local electronic landscape of $\alpha$-MnTe on atomically flat cleaved single crystals. We resolve two distinct electronic regions. In Region A, the chemical potential lies near the valence-band edge and varies by $\sim$100~meV on the nanometer length scale. In Region B, the chemical potential lies near the middle of a wider band gap. We further identify an incommensurate charge modulation with periodicity (2.5$\pm$0.1)$a$, observed exclusively in Region A. Our work establishes that $\alpha$-MnTe can exhibit significant electronic non-uniformity, suggesting that nanoscale characterization is essential for its reliable use in electronic applications.
\end{abstract}

\maketitle

\ptitle{Introduction - altermagnets are cool}
Antiferromagnetic materials are promising candidates for next-generation memory devices due to their ultrafast switching, which can operate in the terahertz range, far surpassing the gigahertz rates of ferromagnets \cite{schlauderer2019Nat}. Unlike ferromagnetic materials, however, conventional antiferromagnets rely on weak relativistic spin-orbit coupling to achieve the spin-dependent transport that is critical for electronic readout of the magnetic state \cite{vzelezny2018NatPhys}. Recently, a new classification of magnetic materials, known as altermagnets, has emerged as a promising platform for novel electronic devices. 
Altermagnets exhibit antiferromagnetic ordering along with strongly spin-split electronic bands, comparable to those of ferromagnets \cite{smejkal2022PRX,jungwirth2026Nat}. This unique combination of properties offers the best of both worlds, potentially combining the ultrafast switching speeds of antiferromagnets with the strong spin-polarization of ferromagnets.

\ptitle{Disorder is an issue for AM}
While initial evidence for altermagnetism emerged from anomalous Hall measurements in leading candidate materials such as RuO$_2$~\cite{feng2022NatElec}, Mn$_5$Si$_3$~\cite{reichlova2024NatComm}, and $\alpha$-MnTe~\cite{gonzalez2023PRL}, subsequent experiments yielded mixed results. The anomalous Hall effect in Mn$_5$Si$_3$ showed pronounced sample-to-sample variation linked to impurity differences~\cite{leiviska2024PRB,han2025PRA}. Furthermore, spin-split bands were not detected in RuO$_2$~\cite{liu2024PRL}, indicating strong sensitivity to sample-dependent factors such as defects~\cite{smolyanyuk2024PRB} and strain~\cite{choi2026NanoConv}. In contrast to RuO$_2$, altermagnetic spin splitting has been observed in $\alpha$-MnTe~\cite{krempasky2024Nat}, yet variations in strain, conductivity, and the N{\'e}el vector direction cause significant variations of the anomalous Hall effect~\cite{liu2025Arxiv, bey2026Arxiv}. These inconsistencies underscore the importance to understand how local material variations modify the electronic structure and the resulting transport signatures of candidate altermagnets. Though recent work imaged monolayer and bilayer MnTe~\cite{DingSciAdv2022, NieNanoLett2023, CuxartAdvFuncMat2026}, these films exhibit a weakly-buckled honeycomb structure with in-plane lattice constants ranging from $a = 4.3$–$4.6$ \AA, deviating significantly from bulk $\alpha$-MnTe. Crucially, the ultrathin film structure is symmetry-incompatible with altermagnetism \cite{CuxartAdvFuncMat2026}. Therefore, high-resolution imaging on pristine bulk $\alpha$-MnTe is imperative to resolve electronic discrepancies relevant to altermagnetic devices.

\ptitle{Here we show}
Here, we use scanning tunneling microscopy and spectroscopy (STM/S) to map electronic disorder on the bulk altermagnetic semiconductor $\alpha$-MnTe. Spectroscopy reveals two distinct electronic environments. In Region A, the chemical potential lies near the valence band and varies continuously by $\sim$100 meV over nanometer length scales. In Region B, the chemical potential is positioned near the center of a wider band gap. We also discover a charge modulation with an incommensurate periodicity of $(2.5 \pm 0.1)a$, where $a$ is the lattice constant. We observe this modulation only in Region A and only at positive sample biases. Our results show that as-grown $\alpha$-MnTe is electronically non-uniform, and bulk signatures should be interpreted in this context.

\begin{figure}
    \centering
    \includegraphics[width=\columnwidth]{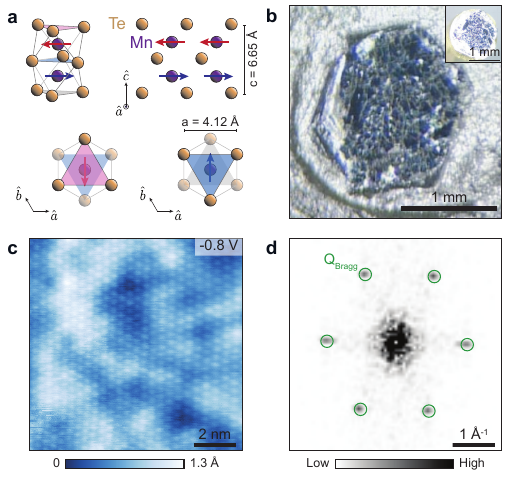}
    \caption{\textbf{$\alpha$-MnTe crystal.}
    (a) Crystal structure of $\alpha$-MnTe (space group $P6_3/mmc$) with lattice constants labeled 
    \cite{moseley2022PRM}. Mn moments, represented by red and blue arrows, are ferromagnetically aligned within each $ab$ plane and antiferromagnetically aligned between adjacent layers.
    (b) Photograph of the cleaved surface of an $\alpha$-MnTe crystal, showing fractured regions of flat surfaces. The inset shows the crystal residue on the cleaving post.
    (c) Constant-current STM topograph of cleaved $\alpha$-MnTe showing atomic-scale contrast. In the NiAs structure, Mn and Te atoms share the same in-plane lattice periodicity, thus the atomic species cannot be unambiguously assigned from topography alone. Sample bias $\Vs=-0.8$~V, setpoint current $\Is=250$~pA.
    (d) FT of the topograph in (c) with green circles marking the Bragg peaks.}
    \label{fig1}
\end{figure}

\ptitle{Crystal structure and negative topo}
Bulk $\alpha$-MnTe crystallizes in the NiAs-type hexagonal structure (space group $P6_3/mmc$). Below the N\'eel temperature ($T_N \approx 307$~K in bulk \cite{ozawa1966PL}), $\alpha$-MnTe orders in an A-type antiferromagnetic ground state where the two Mn sublattices compensate each other, and are connected by a combination of a two-fold spin rotation and a real-space six-fold screw-axis rotation $[C_2 || C_6 \textbf{t}_{1/2}]$ (Fig.\ref{fig1}(a)) \cite{smejkal2022PRXbeyond, amin2024Nat}. We grow $\alpha$-MnTe single crystals from Te flux, and confirm the single phase of our samples by powder X-ray diffraction \cite{moseley2022PRM}. We perform STM/S measurements using electrochemically etched W tips in ultrahigh vacuum at 10~K. We cleave single crystals in situ to obtain flat surfaces (Fig.~\ref{fig1}(b)), on which we achieve, to our knowledge, the first atomically resolved STM topograph on bulk $\alpha$-MnTe (Fig.~\ref{fig1}(c)). The Fourier transform (FT) of the topograph, shown in Fig.~\ref{fig1}(d), displays six Bragg peaks consistent with the hexagonal structure of $\alpha$-MnTe.

\begin{figure*}
    \centering
    \includegraphics[width=1\linewidth]{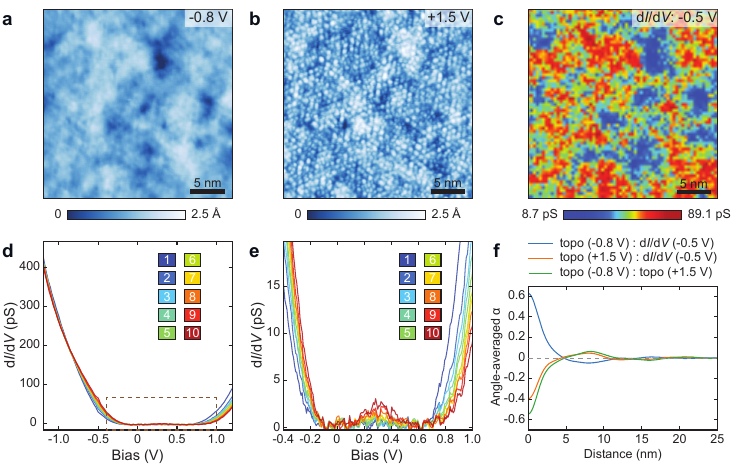}
    \caption{\textbf{Long-range electronic variation.}
    (a,b) Constant-current STM topographs acquired in the same field of view at negative and positive sample bias, respectively. Setpoints: (a) $\Vs=-0.8$~V, $\Is=400$~pA; (b) $\Vs=+1.5$~V, $\Is=200$~pA.
    (c) $dI/dV$ map at $-0.5$~V from a $64\times64$ spectroscopy grid acquired over the same region as (a,b).
    (d) Spatially averaged $dI/dV$ spectra for the 10 bins defined by their value at sample bias $-0.5$~V. Spectra are smoothed along the energy axis using a 32~mV moving-average (boxcar) filter.
    (e) Zoom-in of (d) over the range in the dashed brown box, showing a small in-gap state between 0 and 0.6~V. Spectra were slightly shifted upward to account for a small negative $dI/dV$ contribution. 
    (f) Angle-averaged cross-correlation coefficient $\alpha(r)$ between the $dI/dV$ map at $-0.5$~V [panel (c)] and the low-pass-filtered topographs at $\Vs=-0.8$~V (blue) and $\Vs=+1.5$~V (orange), as well as between the two low-pass-filtered topographs (green). 
    Spectroscopy grid setpoint: $\Vs=-1.3$~V, $\Is=200$~pA, with lock-in modulation (zero-to-peak) $\Ve=6$~mV.
    }\label{fig2}
\end{figure*}

\ptitle{Disorder is electronic}
In addition to the atomic corrugation in Fig.~\ref{fig1}(c), we observe a continuous long-range variation in the topography. To distinguish between structural and electronic sources of this variation, we acquire topographs at different sample biases (Fig.~\ref{fig2}(a,b)). We observe that the contrast of the long-range background is inverted between positive- and negative-bias topographs. To focus on the long-range variation, we apply a low-pass filter to the topographs. The cross-correlation of these filtered images (Fig.~\ref{fig2}(f)) demonstrates anti-correlation with maximum coefficient $\alpha \sim -0.54$. Therefore, the dominant long-range contrast is strongly bias dependent, indicating it primarily reflects electronic variations.

\ptitle{DOS grid explains the disorder}
To uncover the electronic origin of the long-range variation in Fig.~\ref{fig2}(a,b), we measure a grid of $64\times64$ differential conductance ($dI/dV$) spectra covering the same field of view. The $dI/dV$ is proportional to the local density of states (LDOS) as a function of energy. Fig.~\ref{fig2}(c) shows a spatial map of $dI/dV$ at sample bias $-0.5$ V near the top of the valence band. We then sort all $dI/dV$ spectra according to their intensity at $-0.5$ V into ten bins, and present the averaged spectrum for each bin in (Fig.~\ref{fig2}(d)). All spectra exhibit an asymmetric gap of approximately 1~eV, with the chemical potential closer to the valence-band edge. This asymmetry is expected because as-grown $\alpha$-MnTe is slightly hole-doped \cite{gonzalez2023PRL, osumi2024PRB, kriegner2016NatComm} due to native vacancies that shift the chemical potential towards the valence band \cite{ferrer2000PRB}. On the nanoscale, the chemical potential varies by $\sim$100~meV. This variation naturally explains the observed topographic contrast inversion: a local upward shift of the chemical potential increases the LDOS at positive sample biases and decreases it at negative biases. We show that the shift of the chemical potential is spatially correlated with the nanoscale topographic variations in Fig.~\ref{fig2}(f). We also observe in-gap states that appear stronger when the chemical potential is closer to the valence band (Fig.~\ref{fig2}(e)).

\begin{figure*}
    \centering
    \includegraphics[width=1\textwidth]{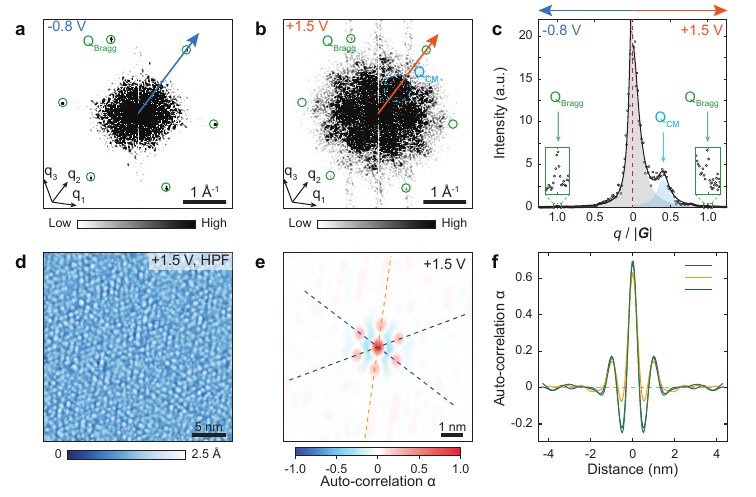}
    \caption{\textbf{Incommensurate charge modulation.}
    (a,b) FT of the topographs shown in Fig.~\ref{fig2}(a,b), respectively. At negative bias $\Vs=-0.8$~V, we observe only the Bragg peaks (green circles). At positive bias $\Vs=+1.5$~V, additional broad peaks ($Q_\text{CM}$, denoted by the dashed blue circle) appear along the Bragg-peak directions $q_1$, $q_2$ and $q_3$, consistent with a disordered charge modulation. We apply a Hanning window before the FT to reduce edge artifacts.
    (c) Left: linecut along $q_2$ of (a) with $\Vs=-0.8$~V. Right: linecut along $q_2$ of (b) with $\Vs=+1.5$~V. Both are averaged over a 10-pixel-wide strip. Shaded areas show the fits to a sum of Lorentzians for the central $q=0$ peak (gray) and the $Q_\text{CM}$ peak (blue) that appears only for the positive-bias linecut. Green boxes show a zoomed-in view around the Bragg peak positions, which were determined by additional Lorentzian components. The horizontal axis is normalized by $|\mathbf{G}|=Q_\mathrm{Bragg}$.
    (d) High-pass-filtered topograph of Fig.~\ref{fig2}(b) with long-range variation removed to enhance the visibility of the charge modulation at $\Vs = 1.5$~V.
    (e) 2D autocorrelation of the filtered topograph in (d).
    (f) Linecuts of the autocorrelation function $\alpha$ along the three directions marked by the dashed lines in (e), averaged over a 10-pixel-wide strip. The profiles illustrate the rapid decay of $\alpha$ from the center, indicating a short correlation length of the charge modulation.}
    \label{fig3}
\end{figure*}

\begin{figure*}
    \centering
    \includegraphics[width=\textwidth]{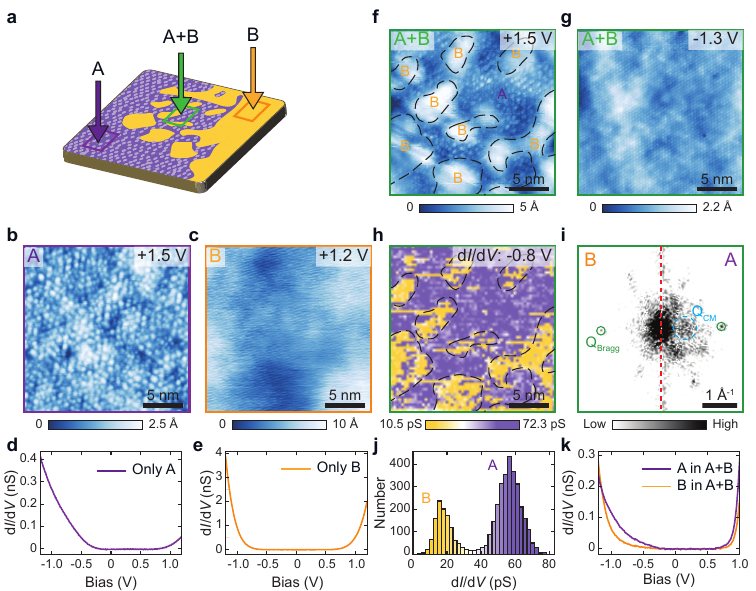}
    \caption{\textbf{Two types of regions with distinct electronic features.} 
    (a) Schematic of MnTe sample surface with different electronic features. Purple (Region A) denotes areas that host disordered charge modulation, orange (Region B) denotes areas without charge modulation, and the green square encompasses a field of view of mixed A and B regions. (b,c) Representative constant-current topographs acquired in A and B regions, respectively.
    (d,e) Spatially averaged $dI/dV$ spectra acquired in the fields of view shown in (b,c), respectively.
    (f) Topograph of a mixed A+B field of view at $\Vs=+1.5$~V; dashed contours trace the region boundaries as a guide to the eye.
    (g) Topograph of the same field of view as (f) at $\Vs=-1.3$~V showing continuous atomic resolution and no charge modulation apparent in either A or B regions at negative bias.
    (h) $dI/dV$ map at $-0.8$~V in the same field of view as (f).
    (i) FT of A (right) and B (left) regions of the field of view (f), masked based on their $dI/dV$ at $-0.8$~V.
    (j) Histogram of $dI/dV$ values at $V_\mathrm{ref}=-0.8$~V from the spectroscopy grid in (h), showing a bimodal distribution associated with A-type and B-type regions.
    (k) Averaged $dI/dV$ spectra for the two populations identified from the bimodal distribution in (j).
    Setpoints: (b) $\Vs=+1.5$~V, $\Is = 200$~pA; (c) $\Vs=+1.2$~V, $\Is = 300$~pA; (d) $\Vs=-1.3$~V, $\Is=200$~pA, $\Ve=6$~mV (zero-to-peak); (e) $\Vs=+1.2$~V, $\Is=300$~pA, $\Ve=6$~mV (zero-to-peak); (f,i) $\Vs=+1.5$~V, $\Is = 100$~pA; (g) $\Vs=-1.3$~V, $\Is = 100$~pA; (h,j,k) $\Vs=-1.3$~V, $\Is=100$~pA; $\Ve=6$~mV (zero-to-peak). Data for Region A and the mixed Region A+B are acquired on one crystal; data for Region B are acquired on a different crystal.}
    \label{fig4}
\end{figure*}

\ptitle{Charge modulation}
The positive bias topograph shown in Fig.~\ref{fig2}(b) exhibits a periodic charge modulation, which is absent at the negative bias (Fig.~\ref{fig2}(a)). Fig.~\ref{fig3}(a,b) show the FT of these topographs, where six broad peaks along the Bragg peak directions appear in the positive-bias FT, but are absent in the negative-bias FT. To quantify this contrast, we take linecuts of both FT maps through one Bragg-peak direction (Fig.~\ref{fig3}(c)), where a peak appears at the wave-vector $Q_\mathrm{CM}$ only for $\Vs=+1.5$~V. The extracted $Q_\text{CM}$ corresponds to a real-space periodicity of $(2.5 \pm 0.1)a$, indicative of an incommensurate modulation. By applying a high-pass filter to the topograph, we isolate the real-space pattern of the charge modulation (Fig.~\ref{fig3}(d)). The modulation is short-ranged and spatially disordered, reminiscent of disordered charge density wave systems \cite{chatterjee2015NatComm}. To quantify the coherence length of these modulations, we calculate a 2D autocorrelation map of the filtered topograph (Fig.~\ref{fig3}(e)) and show that the coefficient decays to nearly zero after roughly 1.5~nm (Fig.~\ref{fig3}(f)).

\ptitle{Wider gap regions}
In different areas of the sample, we observe regions with qualitatively distinct electronic properties (Fig.~\ref{fig4}(a)). We assign Region A to areas that exhibit a periodic charge modulation at positive bias and chemical potential located near the valence-band edge (Fig.~\ref{fig4}(b,d)). In contrast, Region B shows no periodic charge modulation and displays a wider gap with the chemical potential positioned near the gap center (Fig.~\ref{fig4}(c,e)). Although Region B exhibits long-range topographic variation (Fig.~\ref{fig4}(c)), it does not exhibit the contrast inversion between positive and negative biases characteristic of Region A (Fig.~\ref{fig2}). 

\ptitle{Mixed regions}
Beyond the pure areas of Regions A and B, we identify fields of view in which both electronic behaviors coexist. In these mixed areas, the surface exhibits an abrupt nanoscale boundary separating A–like and B–like environments (Fig.~\ref{fig4}(f)). To visualize this fragmented electronic landscape, we measure a grid of $dI/dV$ spectra spanning the same field of view, and histogram the $dI/dV$ values at a reference bias of $V_\text{ref} = -0.8$~V in Fig.~\ref{fig4}(j), revealing a clear bimodal distribution of the two local electronic populations. Mapping this bimodal contrast spatially (Fig.~\ref{fig4}(h)) reveals the sharp nanoscale separation between the two electronic environments: Region A, characterized by a $\sim$1~eV gap with the chemical potential near the valence band, and Region B, characterized by a $\sim$1.5~eV gap with the chemical potential near the center of the gap (Fig.~\ref{fig4}(k)). In Fig.~\ref{fig4}(i), we confirm that the periodic charge modulation appears exclusively in Region A fragments, by masking the positive-bias topograph of Fig.~\ref{fig4}(f) and calculating the FT of A and B fragments separately, shown on the right and left, respectively. Importantly, we confirm continuous atomic resolution across these abrupt electronic changes, by mapping the negative-bias topograph in the same view of view (Fig.~\ref{fig4}(g)). Similar to pure A regions, we observe no charge modulation at negative sample bias throughout this field of view.

\ptitle{Discussion - Region A}
We reveal that $\alpha$-MnTe exhibits two distinct types of disorder: a continuous, nanometer-scale variation of the chemical potential, and an abrupt spatial segregation between two distinct electronic environments (labeled Region A and B in Fig.~\ref{fig4}). Within Region A, the chemical potential skews towards the valence band and varies by $\sim$100 meV over several nanometers (Fig.~\ref{fig2}(c,d)).
This variation illustrates that as-grown $\alpha$-MnTe cannot be assumed to have a spatially-uniform electronic structure, and that bulk signatures should be interpreted with a $\sim$100 meV broadening of the chemical potential in mind. Doping inhomogeneity is a likely origin of this disorder: since as-grown $\alpha$-MnTe is slightly hole-doped due to native vacancies~\cite{ferrer2000PRB}, a spatial inhomogeneity of vacancy concentrations would produce local chemical potential shifts, consistent with our observations. However, we cannot rule out other possible source of disorder such as local strain, which also modifies the electronic structure~\cite{liu2025Arxiv, devaraj2024PRM}. While distinguishing between these scenarios requires further work, our results provide the first microscopic visualization of the electronic inhomogeneity of bulk $\alpha$-MnTe, underlying the macroscopic changes observed for varying strain and conductivity conditions~\cite{liu2025Arxiv, bey2026Arxiv}.

\ptitle{Discussion – Two Regions - Region B and CM}
The $\sim$1~eV gap regions (Region A) are abruptly segregated in space from the $\sim$1.5~eV gap regions (Region B), where the chemical potential lies approximately at mid-gap (Fig.~\ref{fig4}(e,k)). This electronic segregation is further reflected in the presence of the periodic $\sim2.5a$ charge modulation only in Region A (Figs.~\ref{fig2}(b) and \ref{fig4}(b,f)). The correlation between the spectroscopic signatures and the charge modulation indicates that proximity of the valence band to the Fermi level is necessary for the formation of the charge modulation. Moreover, the absence of the modulation at negative sample bias (Figs.~\ref{fig2}(a) and \ref{fig4}(g)) argues against a structural surface reconstruction within Region A and instead supports an electronic origin for this charge modulation.

\ptitle{Conclusions}
In conclusion, we demonstrate that as-grown $\alpha$-MnTe exhibits two electronically distinct regions. In Region A, the chemical potential lies near the valence band and varies by $\sim$100~meV over nanometer length scales. In Region B, the band gap is larger and the chemical potential lies closer to mid-gap. We further discover an incommensurate charge modulation with a period of $\sim2.5a$ that appears only in Region A and only at positive sample bias. Our results demonstrate that a detailed understanding of the nanoscale electronic landscape of $\alpha$-MnTe is essential for its implementation in future electronic devices.

\section*{Acknowledgments}
We thank Rafael Fernandes, Turan Birol, and Luca Buiarelli for useful discussions. G.T.\ was supported by the Israeli Council of Higher Education Quantum Technology Fellowship, and by the Gordon and Betty Moore Foundation's EPiQS Initiative through grant GBMF10215. Work at ORNL was supported by the U.S.\ Department of Energy (DOE), Office of Science, Basic Energy Sciences (BES), Materials Sciences and Engineering Division.


%

\end{document}